\documentclass[10pt,french]{article}
\usepackage{latexsym}
\usepackage{amssymb}
\usepackage{amsmath}
\usepackage{graphicx}
\usepackage[latin1]{inputenc}

\newtheorem{theorem}{Theorem}[section]
\newtheorem{prop}{Proposition}[section]

\newtheorem{lemma}{Lemma}[section]

\newcommand{\R}{\mathbb{R}}             
\newcommand{\N}{\mathbb{N}}             
\newcommand{\C}{\mathbb{C}}             
\renewcommand{\H}{\mathcal{H}}          
\newcommand{\B}{\mathcal{B}}            

\newcommand{\CO}{C_{0}^{\infty}(\R)}    
\newcommand{\comp}{C_{0}^{\infty}}

\newcommand{\half}{\frac{1}{2}}

\newcommand{\Ga}{\Gamma^1}         
\newcommand{\Gb}{\Gamma^2}         

\renewcommand{\l}{\langle}
\renewcommand{\r}{\rangle}
\newcommand{\x}{\l x \r}
\newcommand{\xsi}{\l \xi \r}
\newcommand{\xt}{\frac{x}{t}}

\newcommand{\Pos}{\mathbf{1}_{\R^+}}   
\newcommand{\Neg}{\mathbf{1}_{\R^-}}   
\renewcommand{\d}{\partial}            
\newcommand{\He}{\mathcal{H}_{\textrm{in}}}  
\newcommand{\Hs}{\mathcal{H}_{\textrm{out}}} 
\newcommand{\Pe}{P_{\textrm{in}}}      
\newcommand{\Ps}{P_{\textrm{out}}}     
\newcommand{\We}{W_{\textrm{in}}}      
\newcommand{\Ws}{W_{\textrm{out}}}     
\newcommand{\Js}{J_{\textrm{out}}}     

\newcommand{\Se}{S_{\textrm{in}}}      
\newcommand{\Ss}{S_{\textrm{out}}}     
\newcommand{\Fs}{F_{\textrm{out}}}

\newcommand{\Section}[1]{\section{#1} \setcounter{equation}{0}}

\author{Thierry Daudé \footnote{Department of Mathematics and
    Statistics, McGill University, 805 Sherbrooke South West, Montréal
    QC, H3A 2K6} \, and François Nicoleau
    \footnote{Département de
    Mathématiques, Université de Nantes, 2, rue de la Houssinière, BP
    92208, 44322 Nantes Cedex 03}}
\title{Recovering the mass and the charge of a Reissner-Nordstr\"om
    black hole by an inverse scattering experiment}
\date{}

\begin{document}

\maketitle


\begin{abstract}
  In this paper, we study inverse scattering of massless Dirac fields
  that propagate in the exterior region of a Reissner-Nordstr\"om
  black hole. Using a stationary approach we determine precisely the
  leading terms of the high-energy asymptotic expansion of the
  scattering matrix that, in turn, permit us to recover uniquely the
  mass of the black hole and its charge up to a sign.
\end{abstract}


\Section{Introduction}

Black hole spacetimes are probably among the most fascinating objects
whose existence is predicted by Einstein's general relativity
theory. In the last twenty years, numerous mathematical studies have 
been achieved in order to better understand their
properties. Propagation of fields (wave, Klein-Gordon, Dirac,  
Maxwell) in these peculiar geometries and direct scattering results
have been extensively studied in \cite{Ba1, Ba2, Da, D, DK, Ha1, HaN,
Me2} and used for instance to give rigorous interpretation of the
Hawking effect \cite{Ba3, Ha2, Me3}. Other original phenomena related
to black hole spacetimes are for instance superradiance (the
possibility of extracting energy from a rotating black hole)
\cite{Ba4, FKSY1, FKSY2, FKSY3} or causality violation (the existence
of time machine inside rotating black hole that allows to travel
through time) \cite{Ba5}. Despite the richness of these original
phenomena, a striking feature of black holes spacetimes is their
simplicity. Here we refer to the fact that, whatever might be the
initial configuration of a collapsing stellar body, the resulting
black hole spacetime can be eventually described by ``at most'' three
parameters - its mass, its electric charge and its angular momentum -
an important uniqueness result summarized by the well-known formula:
\emph{Black holes have no hair}. A natural question arises thus:
assume that we are observers living in the exterior region of a black
hole, at rest with respect to it and located far from it (such
observers are called static at infinity and can be typically thought
as a telescope on earth aimed at the black hole), can we in these
conditions measure the defining parameters of the black hole by an 
inverse scattering experiment?  

The result contained in this paper is a first step in this
direction. Here we focus our attention on Reissner-Nordstr\"om black
holes, that is black holes that are spherically symmetric and
electrically charged and thus described by only two parameters: their
mass $M$ and their charge $Q$. We consider massless Dirac fields which
evolve in the exterior region of a Reissner-Nordstr\"om black
hole and we use the direct scattering results, already obtained in
\cite{N1, Me2, Da}, to define the wave operators 
$W^\pm$ and the corresponding scattering operator $S$. The main result
of this paper is then the following. Suppose that the scattering
operator $S$ is known to observers static at infinity (accessible in
theory by physical experiments), then we show that such observers can
recover uniquely the mass $M$ of the black hole as well as its charge
$Q$, up to a sign. Note here that we cannot completely determine the
charge $Q$ of the black hole because of the simplicity of our
model. The propagation of massless Dirac fields 
indeed is only influenced by the geometry (\textit{i.e.} the metric)
of the black hole which depends on $Q^2$ (cf. (\ref{Metric})). The
consideration of massive charged Dirac fields should solve this
undetermination since an explicit term corresponding to the
interaction between the charge of the black hole and that of the field
appear then in the equation (see \cite{Da}). The choice of dealing
with massless Dirac fields is nevertheless meaningful since it is the
simplest model we could study (in particular no long-range terms
appear in the equation) that already provides significant
informations. Moreover, we expect to be able to extend this result to
much more complicated models such as rotating black holes (or Kerr
black holes) for which there already exists a direct scattering theory
\cite{HaN}.   

The strategy we adopt to prove our main result is based on a high-energy 
asymptotic expansion of the scattering operator $S$. Such a technique
was introduced by Enss and Weder in \cite{EW1} and used sucessfully to
recover the potential of multidimensional Schr\"odinger operators
(note that the case of multidimensional Dirac operators in flat
spacetime was treated later by Jung in \cite{Ju}). They showed that
the first term of the high-energy asymptotics is exactly the Radon
transform of the potential they are looking for. Since they work in
dimension greater than two, this Radon transform can be inversed and
the potential thus recovered. In our problem however, due to the
spherical symmetry of the black hole, we are led to study a one 
dimensional Dirac equation and the Radon transform cannot be inversed
in this case. Nevertheless the first term of the high-energy
asymptotics leads to an integral that can fortunately be explicitely
computed and gives exactly the radius of the event horizon, already a
physically relevant information. In order to obtain the mass and the 
charge of the black hole, we must then calculate the second term of
the asymptotics. We follow here the stationary technique introduced by
one of us \cite{Ni1} which is close in spirit to the Isozaki-Kitada
method used in long-range scattering theory \cite{IK}. The basic idea
is to replace the wave operators (and thus the scattering operator) by
explicit Fourier Integral Operators, called modifiers, from which we
are able to compute the high-energy asymptotic expansion readily. The
construction of these modifiers and the precise determination of their
phase and amplitude is the crux of this paper. While this method was
well-known for Schr\"odinger operators and applied successfully to
various situations (see \cite{Ar, Ni1, Ni2, Ni3}), it has required
some substantial modifications when applied to our model due to the
specific properties of Dirac operators, essentially caused by the
matrix-valued nature of the equation. A good starting point to deal
with this difficulty was the paper of G\^atel and Yafaev \cite{GY}
where direct scattering theory of massive Dirac fields in flat
spacetime was studied and modifiers (at fixed energy) were
constructed.


\Section{Reissner-Nordstr\"om black hole and Dirac equation}

In this section, we first briefly describe the geometry of
Reissner-Nordstr\"om black holes. In particular we point out the
main features and properties that are meaningful from scattering
theory viewpoint. We then write down the partial differential
equation that governs the evolution of massless Dirac fields in the
exterior region of Reissner-Nordstr\"om black holes and recall some
known results in direct scattering theory.


\subsection{Reissner-Nordstr\"om black holes}

In Schwarzschild coordinates a Reissner-Nordstr\"om black hole is
described by a four dimensional smooth manifold
\begin{displaymath}
  \mathcal{M} = \mathbb{R}_{t} \times \R^+_r \times
  S_{\omega}^{2},
\end{displaymath}
equipped with the lorentzian metric
\begin{equation} \label{Metric}
  g = F(r)\,dt^{2} - \big(F(r)\big)^{-1} dr^{2} - r^{2}d\omega^{2},
\end{equation}
where
\begin{equation} \label{F}
  F(r) = 1-\frac{2M}{r}+\frac{Q^{2}}{r^{2}},
\end{equation}
and $d\omega^2 = d\theta^{2}+\sin^{2}\theta \, d\varphi^{2}$ is the
euclidean metric on the sphere $S^{2}$. The quantities $M>0$ and $Q
\in \R$ appearing in (\ref{F}) are interpreted as the mass and the
electric charge of the black hole.


The metric (\ref{Metric}) has two types of singularities. Firstly, the
point $\{r = 0\}$ for which the function $F$ is singular. This is a
true singularity or \emph{curvature singularity} \footnote{It means
that certain scalars obtained by contracting the Riemann tensor blow
up when $r \to 0$.}. Secondly, the spheres whose
radii are the roots of $F$ (note that the coefficient of the metric
$g$ involving $F^{-1}$ explodes in this case). The number of these
roots depends on the respective values of the constants $M$ and
$Q$. In this paper we only consider the case $M > |Q|$ for
which the function $F$ has two zeros at the values $r_\pm = M \pm
\sqrt{M^2 - Q^2}$. The spheres $\{r = r_+\}$ and $\{r = r_-\}$ are
called the \emph{exterior} and \emph{interior event horizons} of the
Reissner-Nordstr\"om black hole. Both exterior and interior horizons
are not true singularities in the sense given for $\{r = 0\}$, but
\emph{coordinate singularities}. It turns out that, using appropriate
coordinate systems, these horizons can be understood as regular null
hypersurfaces that can be crossed one way but would require speeds
greater than that of light to be crossed the other way. Hence their
name: event horizons (we refer to \cite{W} for a general introduction
to black hole spacetimes).

Throughout this paper we shall nevertheless use the Schwarzschild
coordinates as the coordinates of our analysis. The reasons are
twofold. First in this coordinate system the coefficients of the
metric (\ref{Metric}) don't depend on the variables $t$ and $\omega$,
reflecting in fact the various symmetries of the black hole. We shall
see in the next subsection \ref{DiracEq} that the Dirac equation will
take a very convenient form, namely an evolution equation (with
respect to $t$) with time-independent Hamiltonian
acting on the spacelike hypersurface $\Sigma = \R_r^+ \times
S^2_\omega$, a nice formalism when dealing with scattering
theory. Second, it corresponds implicitely to the natural notion of
observers static at infinity mentioned in the introduction. Such
observers are located far away from the exterior horizon of the black
hole and live on the integral curves of their velocity $4$-vector
\begin{displaymath}
  U = \frac{1}{F(r)} \frac{\d}{\d t}.
\end{displaymath}
In particular, the time coordinate $t$ is the proper time of such
observers and thus corresponds to their true experience of time. We
refer to \cite{ON} for a more complete discussion of this
notion.. This choice of coordinates appears then quite natural with
the idea of scattering experiments we have in mind.

We point out now a remarkable property of the exterior horizon
when described in Schwarzschild coordinates, property that has
important consequences for scattering theory. From the point of view
of observers static at infinity, the exterior horizon is perceived as
an \emph{asymptotic region} of spacetime. Precisely, it means that the
exterior horizon is never reached in a finite time by incoming and
outgoing null radial geodesics, \textit{i.e} the trajectories followed
by light-rays aimed radially at the black hole or at infinity. These
geodesics are given by the integral curves of $ F(r)^{-1} \frac{\d}{\d
t} \pm \frac{\d}{\d x}$. Along the outgoing geodesics we can express
the time $t$ as a function of $r$ by the formula
\begin{equation} \label{GeodesicsS}
  t(r) = \int_{r_0}^r F(\tau)^{-1} d\tau,
\end{equation}
where $r_0 > r_+$ is fixed. From (\ref{F}) and (\ref{GeodesicsS}), we
see immediately that $t(r) \to -\infty$ when $r \to r_+$. An analogue
formula holds for incoming geodesics.

In consequence we can restrict our attention to the exterior region
$\{r>r_+\}$ of a Reissner-Nordstr\"om black hole and study the inverse
problem for massless Dirac fields there. We just have to keep
in mind that we won't need any boundary conditions on the exterior
event horizon $\{r=r_+\}$ since this horizon is an asymptotic region
spacetime. To make this point even clearer, let us introduce a new
radial coordinate $x$, called the Regge-Wheeler coordinate, which has
the property of straightening the null radial geodesics and will
greatly simplify the later analysis. It is defined implicitely by the
relation 
\begin{equation} \label{RWImplicit}
  \frac{dr}{dx} = F(r),
\end{equation}
and explicitely by
\begin{equation} \label{Regge-Wheeler}
  x = r + \frac{r_+^2}{r_+ - r_-} \log (r-r_+) + \frac{r_-^2}{r_+ -
  r_-} \log (r-r_-).
\end{equation}
In the coordinate system $(t,x, \omega)$, the horizon $\{r=r_+\}$ is
pushed away to $\{x = -\infty\}$ and thanks to (\ref{RWImplicit}), the
metric takes the form
\begin{equation} \label{MetricReggeWheeler}
  g = F(r) (dt^2 - dx^2) -r^2 d\omega^2.
\end{equation}
Observe that the incoming and outgoing null radial geodesics are now
generated by the vector fields $\frac{\partial}{\partial t} \pm
\frac{\partial}{\partial x}$ and take the simple form
\begin{equation} \label{NullGeodesics}
  \gamma^{\pm}(t) = (t, x_0 \pm t, \omega_0), \ \ t \in
  \R,
\end{equation}
where $(x_0, \omega_0) \in \R \times S^2$ are fixed. These are simply
straight lines with velocity $\pm 1$ mimicking, at least in the $t-x$
plane, the situation of a one-dimensional Minkowski spacetime.

From now on we shall work on the manifold  $\B = \R_t \times \R_{x}
\times S^2_{\omega}$ equipped with the metric
(\ref{MetricReggeWheeler}).


\subsection{Dirac equation and direct scattering results} \label{DiracEq}

Scattering theory for Dirac equations on the spacetime $\B$ has been
the object of several papers \cite{N1, Me2, Da} (chronological order)
in the last fifteen years. We  use the following convenient form for a
massless Dirac equation obtained therein. Under Hamiltonian form the
equation eventually reads 
\begin{equation} \label{DiracEquation}
  i \partial_{t} \psi = H \psi,
\end{equation}
where $\psi$ is a $2$-components spinor belonging to the Hilbert space
$\H = L^2(\R \times S^2; \C^2)$ and the Hamiltonian $H$ is given by
\begin{equation} \label{FullDiracOperator}
  H =  \Ga D_x + a(x) D_{S^2},
\end{equation}
where $a(x) = \frac{\sqrt{F(r)}}{r}$, $D_x = -i\d_x$, $D_{S^2}$ denotes
the Dirac operator on $S^2$, \textit{i.e.}
\begin{equation} \label{DiracSphere}
  D_{S^2} = -i \Gb (\partial_{\theta} +
  \frac{\cot{\theta}}{2}) - \frac{i}{\sin{\theta}} \Gamma^3
  \partial_{\varphi},
\end{equation}
and $\Ga, \Gb, \Gamma^3$ appearing in (\ref{FullDiracOperator}) and
(\ref{DiracSphere}) are usual $2 \times 2$ Dirac matrices that satisfy
the anticommutaton relations $\Gamma^i \Gamma^j + \Gamma^j \Gamma^i =
2$Id.  

We can simplify further the expression of the Hamiltonian by using the
spherical symmetry of the equation. The operator $D_{S^2}$ has
compact resolvent and can be diagonalized into an infinite sum
of matrix-valued multiplication operators. The eigenfunctions
associated to $D_{S^2}$ are a generalization of usual spherical
harmonics called \emph{spin-weighted harmonics}. We refer to
I.M. Gel'Fand and Z.Y. Sapiro \cite{GS} for a detailed presentation of
these generalized spherical harmonics.

There exists thus a family of function $F_n^l$ with the
indexes $(l,n)$ running in the set $\mathcal{I} = \big\{ (l,n),
l-|\half| \in \N, l- |n| \in \N \big\}$ which is a Hilbert basis of
$L^2(S^2; \C^2)$ and such that $\forall (l,n) \in
\mathcal{I}$,
\begin{displaymath}
  D_{S^2} F_n^l = -(l+\half) \Gb F_n^l.
\end{displaymath}
The Hilbert space $\H$ can then be decomposed into the infinite direct
sum
\begin{displaymath}
  \H = \bigoplus_{(l,n) \in \mathcal{I}}
  \Big[L^{2}(\R_x; \C^2) \otimes F_{n}^{l} \Big]
  :=\bigoplus_{(l,n) \in \mathcal{I}} \H_{ln},
\end{displaymath}
where $\H_{ln} = L^{2}(\R_x; \C^2) \otimes F_{n}^{l}$ is identified
with $L^2(\R; \C^2)$. We obtain the orthogonal decomposition for
the Hamiltonian $H$
\begin{displaymath}
  H = \bigoplus_{(l,n) \in \mathcal{I}} H^{ln},
\end{displaymath}
with
\begin{equation} \label{DiracOperator}
  H^{ln} := H_{|\H_{ln}} = \Ga D_x + a_l(x) \Gb,
\end{equation}
and $a_l(x) = - a(x) (l + \half)$. The operator $H^{ln}$ is a
selfadjoint operator on $\H_{ln}$ with domain $D(H^{ln}) = H^1(\R;
\C^2)$. Note also that we use the following representation for the
Dirac matrices $\Ga$ and $\Gb$.
\begin{equation} \label{DiracMatrices}
  \Ga = \left( \begin{array}{cc} -1&0 \\0&1 \end{array} \right), \quad
  \quad \Gb = \left( \begin{array}{cc} 0&-1 \\-1&0 \end{array} \right) 
\end{equation}

In this paper it will be enough to restrict our study to a fixed
harmonics. To simplify notation we thus simply write $\H$, $H$ and
$a(x)$ instead of $\H_{ln}$, $H^{ln}$ and $a_l(x)$ respectively.

We summarize now the direct scattering results obtained in
\cite{N1, Me2, Da}. We define $H_0 = \Ga D_x$ as the free
Hamiltonian. It is clearly a selfadjoint operator on $\H$ with domain
$D(H_0) = H^1(\R; \C^2)$. The main results concerning the spectral
properties of the pair $(H, H_0)$ are the following. The Hamiltonian
$H$ has no eigenvalue and no singular continuous spectrum,
\textit{i.e.} 
\begin{equation} \label{AbsenceEigenvalue}
  \sigma_{pp} (H) = \emptyset, \quad \sigma_{sc}(H) = \emptyset, 
\end{equation}
and the standard wave operators
\begin{equation} \label{WaveOperator}
  W^\pm = s-\lim_{t \to \pm \infty} e^{itH}e^{-itH_0},
\end{equation}
exist and are complete on $\H$. Let us make a few comments on these
results. It follows directly from (\ref{AbsenceEigenvalue})
and (\ref{WaveOperator}) that $H$ has purely absolutely continuous
spectrum and thus all states scatter in the asymptotic regions of
spacetime where they obey a simpler equation given by the comparison
dynamics $e^{itH_0}$. Notice that this comparison dynamics is
particularly simple for massless Dirac fields. Due to the diagonal
form of the matrix $\Ga$, it becomes simply a system of transport
equations with velocity $+1$ for the second component of the spinor
and velocity $-1$ for the first. At last, observe that the potential
$a$ has the following asymptotics as $x \to \pm \infty$:  
\begin{displaymath}
  \exists \kappa>0, \quad |a(x)| = O(e^{\kappa x}), \quad x \to
  -\infty,
\end{displaymath}
\begin{displaymath}
  |a(x)| = O(|x|^{-1}), \quad x \to +\infty.
\end{displaymath}
Eventhough $a$ is apparently a long-range potential having Coulomb
decay at $+\infty$, we don't need to modify the comparison dynamics
in the definition of the wave operators. We refer to \cite{Da},
Section 7 for a proof of this point.

The scattering operator $S$ is defined by
\begin{equation} \label{ScatteringOperator}
  S = (W^+)^* W^-.
\end{equation}
We can simplify the operator $S$ by considering incoming and
outgoing initial data separately\footnote{The terms incoming and
outgoing data have to be understood here in reference to the black
hole, \textit{i.e.} as the data whose associated solutions propagate,
when $t$ increases, in direction of the black hole (incoming data)
or escape from it (outgoing data)}. In order to define these
subspaces of initial data precisely, let us introduce the asymptotic
velocity operators
\begin{displaymath}
  P_0^\pm = s-C_\infty-\lim_{t \to \pm \infty} e^{itH_0} \xt
  e^{-itH_0},
\end{displaymath}
and
\begin{displaymath}
  P^\pm = s-C_\infty-\lim_{t \to \pm \infty} e^{itH} \xt
  e^{-itH}.
\end{displaymath}
We refer to \cite{DG} for a detailed presentation of these operators
and their usefulness in scattering theory. By a direct calculation,
we get easily $P_0^\pm = \Ga$ while it was shown  in \cite{Da} that
the operators $P^\pm$ exist, have also spectra equal to $\{-1, +1\}$
and satisfy the intertwining relations
\begin{equation} \label{Intertwining}
  P^\pm W^\pm = W^\pm \Ga.
\end{equation}
Let us denote $\Pe = \mathbf{1}_{\R^-}(\Ga)$ and $\Ps \mathbf{1}_{\R^+}(\Ga)$ the projections onto the negative and positive
spectral subspaces of the ``free'' asymptotic velocity $\Ga$. Then we
define $\He^0 = \Pe \H$ and $\Hs^0 = \Ps \H$ and refer to these spaces
as incoming and outgoing initial data for the free
dynamics\footnote{Observe that $\He^0$ (resp. $\Hs^0$) are simply the
spinors with nonvanishing first component (resp. second component)
only}. In a similar way, we define $\He^\pm = \mathbf{1}_{\R^-}(P^\pm)
\H$ and $\Hs^\pm = \mathbf{1}_{\R^+}(P^\pm) \H$ the spaces of incoming
and outgoing initial data for the full dynamics when $t \to
\pm\infty$. In our particular case it turns out that we have
equalities between some of these spaces. Precisely we have
\begin{lemma} \label{InOut}
  With the same notations as above, we have 
  \begin{eqnarray*}
    \He^+ = \He^- & =: & \He, \\ 
    \Hs^+ = \Hs^- & =: & \Hs.
  \end{eqnarray*}
\end{lemma}
\textbf{Proof}: We only prove the case ``out'' since the proof for
``in'' is similar. We shall use the following characterization of the
asymptotic velocity operators obtained in \cite{Da}
\begin{equation} \label{Char}
  P^\pm = s-C^\infty-\lim_{t \to \pm \infty} e^{itH} \Ga e^{-itH}.
\end{equation}
We compute $\Neg(P^-) \Pos(P^+)$. By (\ref{Char}) we get
\begin{displaymath}
  \Neg(P^-) \Pos(P^+) = s-\lim_{t \to \pm \infty} e^{-itH}
  \Neg(\Ga) e^{2itH} \Pos(\Ga) e^{-itH}.
\end{displaymath}
Since $\Ga H = \tilde{H} \Ga$ with $\tilde{H} = \Ga D_x - a(x) \Gb$,
we have
\begin{eqnarray*}
  \Neg(P^-) \Pos(P^+) & = & s-\lim_{t \to \pm \infty} e^{-itH}
  e^{2it\tilde{H}} \Neg(\Ga) \Pos(\Ga) e^{-itH}, \\
                      & = & 0.
\end{eqnarray*}
Hence $\Pos(P^+) = \big( \Neg(P^-) + \Pos(P^-) \big) \Pos(P^+) \Pos(P^-) \Pos(P^+)$ from which we deduce immediately that $\Hs^+
\subset \Hs^-$. Analogously we can prove $\Hs^- \subset \Hs^+$ and
thus the equality $\Hs^+ = \Hs^-$ holds. \\ $\diamondsuit$ \\

Thanks to the intertwining relations (\ref{Intertwining}) and Lemma
\ref{InOut} we see that the wave operators $W^\pm$ are partial
isometries from $\H^0_{\textrm{in/out}}$ into
$\H_{\textrm{in/out}}$. We define
\begin{displaymath}
  \Ws^{\pm} = s-\lim_{t \to \pm\infty} e^{itH} e^{-itH_0} \Ps,
\end{displaymath}
\begin{displaymath}
  \We^{\pm} = s-\lim_{t \to \pm\infty} e^{itH} e^{-itH_0} \Pe,
\end{displaymath}
the corresponding outgoing and incoming wave operators. Remark here
that by the relations $H_0 \mathbf{1}_{\R^\pm}(\Ga) = \pm D_x
\mathbf{1}_{\R^\pm}(\Ga)$ the Hamiltonian $H_0$ becomes ``scalar''
when projected on $\H_{\textrm{in/out}}^0$. Hence we can simplify the
expression of the outgoing and incoming wave operators by
\begin{displaymath}
  \Ws^{\pm} = s-\lim_{t \to \pm\infty} e^{itH} e^{-itD_x} \Ps,
\end{displaymath}
\begin{displaymath}
  \We^{\pm} = s-\lim_{t \to \pm\infty} e^{itH} e^{itD_x} \Pe.
\end{displaymath}
This minor remark will turn out to be important in the next
section. Now from the definition (\ref{ScatteringOperator}) of the
scattering operator, we deduce that $S$ leaves invariant the incoming
and outgoing subspaces $\H_{\textrm{in/out}}^0$ and thus can be
decomposed into the direct sum
\begin{equation} \label{DecomposedS}
  S = \Se \oplus \Ss,
\end{equation}
where
\begin{equation} \label{ReducedS}
  \Se = (\We^+)^* \We^-, \quad \Ss = (\Ws^+)^* \Ws^-.
\end{equation}
Clearly $\Se$ and $\Ss$ are isometries on $\He^0$ and $\Hs^0$.


\Section{The inverse problem at high energy}


\subsection{The main result and the strategy of the proof}

We are now ready to study the inverse problem for massless Dirac
fields satisfying (\ref{DiracEquation}). The question we adress is:
can we determine the mass $M$ and the charge $Q$ of the black
hole from the knowledge of the scattering operator $S$? The main
result of this paper is summarized in the following Theorem
\begin{theorem} \label{MainThm}
  Assume that the scattering matrix $\Ss$ (or $\Se$) is known. Then the
  corresponding potential $a(x)$ can be entirely determined. Moreover
  the mass $M$ and the square of the charge $Q^2$ of the black hole
  can be recovered by the formulas 
  \begin{equation} \label{Mass}
    M = \lim_{x \to +\infty} \frac{x - x^3 a^2(x)}{2},
  \end{equation}
  \begin{equation} \label{Charge}
    Q^2 = \lim_{x \to +\infty} (x^4 a^2(x) - x^2 + 2M x).
  \end{equation}
\end{theorem}

The strategy we adopt to prove this Theorem is based on a high-energy
asymptotic expansion of $\Ss$, a well-known technique initially
developed in the case of multidimensional Schr\"odinger operators by
Enss, Weder \cite{EW1}. This method can be used to study Hamiltonians
with electric and magnetic potentials \cite{Ar}, the Dirac equation
\cite{Ju} and Stark Hamiltonians \cite{We,Ni3}. 

Precisely, we consider $\psi_1, \psi_2 \in \Hs^0$ such
that $\hat{\psi_1}, \hat{\psi_2} \in \comp(\R ; \C^4)$
and we define the map
\begin{equation} \label{F(lambda)}
  \Fs(\lambda) = < \Ss e^{i\lambda x} \psi_1, e^{i\lambda x} \psi_2
  >,
\end{equation}
for $\lambda \in \R$. Following the ideas of \cite{Ni1, Ni2}, we shall
obtain an asymptotic expansion of $S(\lambda)$ when $\lambda \to
\infty$ with the exact expression of the first three leading
terms. Exactly we get the following \emph{reconstruction formula}:
\begin{theorem}[Reconstruction formula] \label{ReconstructionFormula}
  Let $\psi_1, \psi_2 \in \Hs^0$ such that $\hat{\psi_1}, \hat{\psi_2}
  \in \comp(\R ; \C^2)$. Then for $\lambda$ large, we obtain
  \begin{equation} \label{RF}
    \Fs(\lambda) = <\psi_1, \psi_2> + \frac{1}{\lambda} <\psi_1, L_1(x,D_x)
    \psi_2> + \frac{1}{\lambda^2} <\psi_1, L_2(x,D_x) \psi_2> + \
    O(\lambda^{-3}),
  \end{equation}
  where $L_j(x,D_x)$ are differential operators given by
  \begin{displaymath}
    L_1(x,D_x) = L_1 = \frac{i(l+\half)^2}{2r_+},
  \end{displaymath}
  \begin{displaymath}
    L_2(x,D_x) = \frac{a_l^2(x)}{2} + \frac{(l+\half)^4}{8r_+^2} - \frac{i
    (l+\half)^2 D_x}{2 r_+},
  \end{displaymath}
  and $r_+ = M + \sqrt{M^2 - Q^2}$.
\end{theorem}
Let us admit temporarily this result and use it to prove the main
theorem. If we assume that the scattering matrix $\Ss$ is known, then
we can use inductively the high-energy expansion (\ref{RF}) to recover
first, the radius $r_+$ of the exterior event horizon (term of order
$\lambda^{-1}$) and second, the potential $a(x)$. Indeed, the term of
order $\lambda^{-2}$ (since $r_+$ is already determined) gives the
quantity  
\begin{equation} \label{Term2}
  < \frac{a_l^2(x)}{2} \psi_1, \psi_2>.
\end{equation}
Since (\ref{Term2}) can be determined for any $\psi_1, \psi_2$ in a
dense subset in $\H$, we recover completely the potential $a_l(x)$ and
thus $a(x)$. Now the formulae for the mass (\ref{Mass}) and for the
charge (\ref{Charge}) follow directly from the definition of $a(x)$  
\begin{displaymath}
  a^2(x) = \frac{1 - \frac{2M}{r} + \frac{Q^2}{r^2}}{r^2},
\end{displaymath}
and the fact that
\begin{displaymath}
  r(x) \sim  x \ ,  x \to +\infty
\end{displaymath}
an immediate consequence of (\ref{Regge-Wheeler}). This ends the proof
of our main Theorem and proves that observers static at infinity who
can measure the scattering matrix are able to recover the mass of the
black hole and its charge, up to a sign. Note that the first term of
the expansion (\ref{RF}) permits us to determine the radius $r_+$ of
the exterior event horizon, an information of physical relevance.    

The main step in the proof of Theorem \ref{MainThm} is
clearly the reconstruction formula \ref{RF} stated above. In order to
prove it, we first rewrite $\Fs(\lambda)$ as follows 
\begin{eqnarray}
  \Fs(\lambda) & = & < \Ss e^{i\lambda x} \psi_1, e^{i\lambda x}
  \psi_2>, \nonumber \\
             & = & < \Ws^- e^{i\lambda x} \psi_1, \Ws^+
  e^{i\lambda x} \psi_2>, \nonumber \\
             & = & <\Ws^-(\lambda) \psi_1, \Ws^+(\lambda)
  \psi_2>, \label{ExpressionF}
\end{eqnarray}
where
\begin{displaymath}
  \Ws^\pm(\lambda) = s-\lim_{t \to \pm \infty} e^{itH(\lambda)}
  e^{-it(D_x + \lambda)} \Ps,
\end{displaymath}
and
\begin{displaymath}
  H(\lambda) = \Ga (D_x + \lambda) + a(x) \Gb.
\end{displaymath}
In order to obtain an asymptotic expansion of the operators
$\Ws^\pm(\lambda)$, we follow the procedure exposed in \cite{Ni1, Ni2},
procedure inspired by the well-known Isozaki-Kitada method \cite{IK}
developed in the setting of long-range stationary scattering
theory. It consists simply in replacing the wave operators
$\Ws^\pm(\lambda)$ by ``well-chosen'' energy modifiers
$\Js^\pm(\lambda)$, defined as Fourier Integral Operators (FIO) with
explicit phase and amplitude. Well-chosen here means practically that
we look for $\Js^\pm(\lambda)$ satisfying for $|\lambda|$ large
\begin{equation} \label{Property1}
  \Ws^\pm(\lambda) \psi = \lim_{t \to \pm \infty} e^{itH(\lambda)}
  \Js^\pm(\lambda) e^{-it(D_x + \lambda)} \psi,
\end{equation}
and
\begin{equation} \label{Property2}
  \| (\Ws^\pm(\lambda) - \Js^\pm(\lambda)) \psi \| =  O(\lambda^{-3}),
\end{equation}
for a fixed $\psi \in \Hs^0$ such that $\hat{\psi} \in \comp(\R ; \C^2)$. In particular if we manage to construct
$\Js^\pm(\lambda)$ satisfying (\ref{Property2}) then we obtain by
(\ref{ExpressionF})
\begin{displaymath}
  \Fs(\lambda) = \ <\Js^-(\lambda) \psi_1, \Js^+(\lambda) \psi_2>
  + \ O(\lambda^{-3}),
\end{displaymath}
from which we can calculate the first terms of the asymptotics easily.

The rest of this paper is devoted to the construction of the modifiers
$\Js^\pm(\lambda)$ and to the proof of Theorem
\ref{ReconstructionFormula}. For simplicity of notations we omit in
the next subsections the underscript ``out'' when naming the wave
operators $\Ws^\pm$ and modifiers $\Js^\pm$.


\subsection{Construction of the modifier}

Let us give a hint on how to construct the modifiers
$J^\pm(\lambda)$ a priori defined as an FIO with ``scalar'' phase
$\varphi^\pm(x,\xi,\lambda)$ and ``matrix-valued'' amplitude
$p^\pm(x,\xi,\lambda)$. If we assume that (\ref{Property1}) is true
then we can write
\begin{displaymath}
  (W^\pm(\lambda) - J^\pm(\lambda)) \psi = i \int_0^{\pm
  \infty} e^{itH(\lambda)} C^\pm(\lambda) e^{-it (D_x + \lambda)}
  \psi dt,
\end{displaymath}
where
\begin{equation} \label{C}
  C^\pm(\lambda) := H(\lambda) J^\pm(\lambda) -
  J^\pm(\lambda) (D_x + \lambda),
\end{equation}
is an FIO with phase $\varphi^\pm(x,\xi,\lambda)$ and amplitude
$c^\pm(x,\xi,\lambda)$. Hence we get the simple estimate
\begin{equation} \label{Property3}
  \| (W^\pm(\lambda) - J^\pm(\lambda)) \psi \| \leq \int_0^{\pm
  \infty} \| C^\pm(\lambda) \ e^{-it D_x}  \psi  \| dt.
\end{equation}
In order that (\ref{Property2}) be true it is then clear that the FIO
$C(\lambda)$ has to be ``small'' in some sense. Precisely we shall
need that the amplitude $c^\pm(x,\xi,\lambda)$ be short-range in the
variable $x$ and of order $\lambda^{-3}$ in the variable $\lambda$.


\subsubsection{Construction of the modifiers at fixed energy}

We first look at the problem at fixed energy (\emph{i.e.} we take
$\lambda = 0$ in the previous formulae). Hence we aim to construct
modifiers $J^\pm$ with scalar phase $\varphi^\pm(x,\xi)$ and
matrix-valued amplitude $p^\pm(x,\xi)$ such that the amplitude
$c^\pm(x,\xi)$ of the operator $C^\pm = H J^\pm - J^\pm D_x$
be short-range in $x$. We follow here the treatment given by Gâtel and
Yafaev in \cite{GY} where a similar problem for massive Dirac
Hamiltonians in Minkowski spacetime was considered.

The operator $C^\pm$ is clearly an FIO with phase
$\varphi^\pm(x,\xi)$ and amplitude
\begin{equation} \label{AmplitudeC}
  c^\pm(x,\xi) = B^\pm(x,\xi) p^\pm(x,\xi) - i\Ga \d_x p^\pm(x,\xi),
\end{equation}
where
\begin{equation} \label{B}
  B^\pm(x,\xi) = \Ga \d_x \varphi^\pm(x,\xi) + a(x) \Gb - \xi.
\end{equation}
As usual we look for a phase $\varphi^\pm$ close to $x \xi$ and an
amplitude $p^\pm$ close to $1$. So the term $\d_x p^\pm$ in (\ref{AmplitudeC})
should be short-range et can be neglected. Moreover if $p^\pm$ was exactly
$1$ we would have to determine $\varphi^\pm$ so that
\begin{equation} \label{B=0}
  B^\pm = 0.
\end{equation}
A direct calculation of $B^\pm=0$ leads to a matrix-valued phase
$\varphi^\pm$ whereas we look for a scalar one. But if $\varphi^\pm$
satisfies (\ref{B=0}) then it should also satisfy the equation $B^2 =
0$. Using the anticommutation properties of the Dirac matrices we have 
\begin{equation} \label{BE}
  (B^\pm)^2(x,\xi) = (\d_x \varphi^\pm)^2 + a^2(x) - \xi^2 - 2 \xi
  B^\pm(x,\xi) = 0.
\end{equation}
Using (\ref{B=0}) again, we eventually get the scalar equation
\begin{equation} \label{B2}
  r^\pm(x,\xi) := (\d_x \varphi^\pm)^2 + a^2(x) - \xi^2 = 0,
\end{equation}
in fact an eikonal equation of Schr\"odinger type. Let us set
$\varphi^\pm(x,\xi) = x \xi + \phi^\pm(x,\xi)$ where $\phi^\pm(x,\xi)$
should be a priori small in the variable $x$. Then we must solve $2\xi
\d_x \phi^\pm + (\d_x \phi^\pm)^2 + a^2(x) = 0$. Neglecting
$(\d_x\phi^\pm)^2$ in this last equation, we obtain
\begin{equation} \label{Eikonal}
  2\xi \d_x \phi^\pm + a^2(x) = 0.
\end{equation}

\par\noindent
For $\xi \not=0$, we get the following two solutions of
(\ref{Eikonal}) by the formulae 
\begin{equation} \label{Phase}
  \phi^\pm(x,\xi) = \frac{1}{2\xi} \int_0^{\pm \infty} a^2(x+s) ds,
\end{equation}
and with this choice, we obtain for $\xi \not=0$,
\begin{equation} \label{Rest}
  r^\pm(x,\xi) = \frac{1}{4\xi^2}  \ a^4(x).
\end{equation}

\vspace{0,3cm}
In our derivation of the phase, it is important
to keep in mind that we didn't find an approximate solution of
(\ref{B=0}) but instead of (\ref{BE}). Therefore we cannot expect to
take $p^\pm = 1$ as a first approximation and we have to work a bit
more. In order to construct the amplitude $p^\pm$ we follow here
particularly closely \cite{GY}.

So we look for $p^\pm$ such that $B^\pm p^\pm$ is as small as
possible. What we know is that $(B^\pm)^2$ is already small thanks to
our choice of $\varphi^\pm$. Hence let's try to find a relation
between $B^\pm$ and $(B^\pm)^2$. Observe first that $B^\pm$ have
the following expression on $\Gamma^\pm$
\begin{equation} \label{B1}
  B^\pm(x,\xi) = B(x,\xi) = B_0(x,\xi) + 2 \xi K(x,\xi),
\end{equation}
where
\begin{eqnarray}
  B_0(x,\xi) & = & -2 \xi \Pe, \label{B0} \\
  K(x,\xi) & = & \frac{1}{2\xi} \big( -\frac{a^2(x)}{2\xi} \Ga +
  a(x)\Gb \big). \label{K}
\end{eqnarray}
In particular, $B^\pm$ don't depend on $\pm$ in $\Gamma^\pm$. Taking
the square of (\ref{B1}) we get
\begin{equation} \label{B3}
  B^2 = B_0^2 + 2\xi B_0 K + 2\xi K B
\end{equation}
But from (\ref{B0}) we see that $B^2_0 = -2\xi B_0$. Whence (\ref{B3})
becomes
\begin{equation} \label{B4}
  B^2 = -2\xi B_0 (1-K) + 2\xi K B.
\end{equation}
If we add $2\xi B$ to both sides of the equality in (\ref{B4}) then
we obtain by (\ref{BE}),
\begin{equation}
  r(x,\xi) := B^2 + 2\xi B = -2\xi B_0 (1-K) + 2\xi (1+K) B,
\end{equation}
or equivalently, if we isolate $B$ on the left hand side
\begin{equation}
  2\xi (1+K) B = r(x,\xi) - 4 \xi^2 \Pe (1-K).
\end{equation}
It follows immediately from the definition of $K$ that $(1 \pm K)$ are
inversible for $\xi \in X:= \{ \ \xi \in \R  \ :\ \mid \xi \mid \geq R
\ \}$, $R \gg 1$. In consequence we can write for  $\xi \in X$
\begin{equation} \label{B5}
  B (1-K)^{-1} = \frac{1}{2\xi} (1+K)^{-1} r(x,\xi) (1-K)^{-1} -2\xi
  \Pe.
\end{equation}
The first term in the right hand side of (\ref{B5}) is small thanks to
our choice of phase but the second one is not. We choose $p^\pm$ in
such a way that it cancels this term. According to (\ref{B5}), a
natural choice for $p^\pm$ is thus
\begin{equation} \label{p}
  p^\pm(x,\xi) = p(x,\xi) = (1-K)^{-1} \Ps,
\end{equation}
for which we have
\begin{equation} \label{q}
  q(x,\xi) := B(x,\xi) p(x,\xi) = \frac{1}{2\xi} (1+K)^{-1} r(x,\xi)
  (1-K)^{-1}.
\end{equation}
Note that $p$ defined by (\ref{p}) is independent of $\pm$.

\vspace{0,2cm}
\par\noindent
Let us summarize the situation at this stage. For $\xi \not= 0$, we
have defined the phase $\varphi^\pm (x,\xi) = x\xi + \phi^\pm(x,\xi) $
by (\ref{Phase}) and for $\xi \in X$, the amplitude $p$ is given by
(\ref{p}). Directly from the definitions, the following estimates
hold.

\begin{lemma}[Estimates on the phase and the amplitude]
\begin{equation} \label{EstPhase1}
  \forall (x,\xi) \in \R \times X, \forall \alpha, \beta \in \N,
  \quad |\d_x^\alpha \d_\xi^\beta \varphi^\pm(x,\xi)| \leq C_{\alpha\beta}
  \ (1+ \mid x\mid + \mid \xi \mid)^{(2-\alpha-\beta)_+}.
\end{equation}
\begin{equation} \label{EstPhase2}
  \forall (x,\xi) \in \R \times X,
  \quad |\d_{x,\xi}^2 (\varphi^\pm(x,\xi)- x\xi)| \leq \frac{C}{R^2}
\end{equation}
\begin{equation} \label{EstPhase3}
  \forall (x,\xi) \in \R^{\pm} \times X, \forall \alpha, \beta \in \N,,
\quad |\d_x^\alpha \d_\xi^\beta \phi^\pm(x,\xi)| \leq C_{\alpha\beta}
  \ \x^{-1-\alpha} \xsi^{-1-\beta}.
\end{equation}

\begin{equation} \label{EstK}
  \forall (x,\xi) \in \R \times X, \forall \alpha,
  \beta \in \N, \quad |\d_x^\alpha \d_\xi^\beta K(x,\xi)| \leq C_{\alpha\beta} \
  \x^{-1-\alpha} \xsi^{-1-\beta}.
\end{equation}
\begin{equation} \label{EstP}
  \forall (x,\xi) \in \R \times X, \forall \alpha,
  \beta \in \N, \quad |\d_x^\alpha \d_\xi^\beta (p(x,\xi) - \Ps)| \leq
  C_{\alpha\beta} \ \x^{-1-\alpha} \xsi^{-1-\beta}.
\end{equation}
\begin{equation} \label{EstR}
  \forall (x,\xi) \in \R \times  X, \forall \alpha,
  \beta \in \N, \quad |\d_x^\alpha \d_\xi^\beta r(x,\xi)| \leq
  C_{\alpha\beta} \  \x^{-4-\alpha} \xsi^{-2-\beta}.
\end{equation}
\begin{equation} \label{EstQ}
  \forall (x,\xi) \in \R \times  X, \forall \alpha,
  \beta \in \N, \quad |\d_x^\alpha \d_\xi^\beta q(x,\xi)| \leq
  C_{\alpha\beta} \ \x^{-4-\alpha} \xsi^{-3-\beta}.
\end{equation}
\end{lemma}

Thanks to (\ref{EstPhase1}), (\ref{EstPhase2}) and (\ref{EstP}), for
$R$ large enough, we can define precisely our modifier $J^{\pm}$ as a
bounded operator on $\H$  (\cite{Ro}, Corollary IV.22): let $\theta
\in C^\infty(\R)$ be a cutoff function such that $\theta(\xi) = 0$ if
$|\xi| \leq \frac{1}{2}$ and $\theta(\xi) = 1$ if $|\xi| \geq 1$. For
$R$ large enough, $J^{\pm}$ is the Fourier Integral Operator with
phase $\varphi^{\pm} (x, \xi)$ and amplitude 
\begin{equation} \label{Truncatedamplitude} 
  P(x,\xi) = p(x,\xi)\ \theta(\frac{\xi}{R})
\end{equation}

\par
\noindent
Moreover, we have
\begin{prop} \label{Mod}
 For any $\psi \in \Hs^0$ such that $\hat{\psi} \in
  \comp(X; \C^2)$, we have
  \begin{equation}
    W^\pm \psi = \lim_{t \to \pm \infty} e^{itH} J^\pm e^{-itD_x}
    \psi.
  \end{equation}
\end{prop}
\textbf{Proof}: We only consider the case $(+)$. It is clear that it
suffices to show 
\begin{equation} \label{Perturb}
  \lim_{t \rightarrow + \infty} \ \left( J^+ - \theta (\frac{D_x}{R})
  \Ps \right) \ e^{-itD_x} \psi \ =\ 0. 
\end{equation}
Let $\eta \in C^\infty(\R)$ such that $\eta (x)=1$ if $x \geq 1$ and
$\eta (x)=0$ if $x \leq 0$. We write $e^{-itD_x} \psi$ as :
\begin{equation} \label{decoup}
e^{-itD_x} \psi \ = \ \eta (x)\  e^{-itD_x} \psi + (1-\eta(x)) \
e^{-itD_x} \psi 
\end{equation}
Clearly, $(1-\eta(x)) \ e^{-itD_x} \psi =  e^{-itD_x} \ (1-\eta(x+t))
\ \psi \rightarrow 0$ on $\H$. 
\vspace{0,1cm}
\par\noindent
Now, using (\ref{EstPhase1}), (\ref{EstPhase3}), (\ref{EstP}) and the standard
pseudodifferential calculus,
we see that \\
${\displaystyle{ T:= \left(J^+ -\ \theta (\frac{D_x}{R}) \Ps \right)
    \eta (x)}}$ is a $\Psi$do  with symbol $k^+(x,\xi)$ which
satisfies 
\begin{equation} \label{compact}
\forall (x,\xi) \in \R \times  \R, \forall \alpha, \beta \in \N, \quad
  |\d_x^\alpha \d_\xi^\beta k^+(x,\xi)| \leq 
  C_{\alpha\beta} \  \x^{-1-\alpha} \xsi^{-1-\beta}.
  \end{equation}
Then $T$ is a compact operator on $\H$ and since
$e^{-itD_x} \psi \rightarrow 0$ weakly,  Proposition \ref{Mod} is
proved. \\ $\diamondsuit$ \\


\subsubsection{Construction of the modifiers at high energy}

Let's come back to the construction of $J^\pm(\lambda)$ the modifiers
at high energy. Recall that we look for modifiers that satisfy
(\ref{Property1}) and (\ref{Property2}). Proposition \ref{Mod}
suggests to construct $J^\pm(\lambda)$ close to $e^{-i\lambda x} J^\pm e^{i\lambda x}$.

\par\noindent
Clearly, $e^{-i\lambda x} J^\pm e^{i\lambda x}$ are the  FIOs with phases $\varphi^\pm(x,\xi,\lambda) = x\xi +
\Big( \frac{1}{2(\xi + \lambda)} \int_0^{\pm \infty} a^2(x+s) ds \Big)$ and amplitude
$P(x,\xi + \lambda)$. Since the modifiers $J^\pm(\lambda)$ will be applied on functions $\psi$
having compact support in the Fourier variable, the cut-off in the $\xi$ variable in the definition of
$P(x,\xi +\lambda)$ disappears for $\lambda$ sufficiently large enough.

\vspace{0,2cm}
\par\noindent
Thus, if we take exactly $J^\pm(\lambda)= e^{-i\lambda x} J^\pm
e^{i\lambda x}$ as the modifiers, according to (\ref{AmplitudeC}),
for $\xi$ in a compact set and $\lambda \gg 1$,
the amplitude of the operator $C^\pm(\lambda)$ becomes
\begin{displaymath}
  c^\pm (x,\xi,\lambda) = c(x,\xi,\lambda) = B(x, \xi + \lambda)
  p(x,\xi+ \lambda) -i \Ga  \d_x p(x,\xi + \lambda).
\end{displaymath}
We want the amplitude $c(x,\xi,\lambda)$ to be of order $O(\x^{-2}
\lambda^{-3})$. By (\ref{EstQ}) the first term $B(x, \xi +
\lambda) p(x,\xi+ \lambda)$ has order $O(\x^{-4} \lambda^{-3})$.
By (\ref{EstP}) the second term $\d_x p(x, \xi + \lambda)$ has
order $O(\x^{-2} \lambda^{-1})$ which is not a sufficient decay in
$\lambda$ for our purpose.

\vspace{0,2cm}
\par\noindent
So, we have to work a bit more. We look for the modifiers $J^\pm(\lambda)$, with $\lambda$ large enough, as FIOs
with phases $\varphi^\pm(x,\xi,\lambda)$ and with a new amplitude $P(x,\xi,\lambda)$ under the form
\begin{displaymath}
  P(x,\xi,\lambda) = \left( p(x,\xi+\lambda) + \frac{1}{\lambda^2} \Big(
  \Pe k_1(x,\xi) + p(x,\xi+\lambda) l_1(x,\xi) \Big) +
  \frac{1}{\lambda^3} \Pe k_2(x,\xi) \right) \ g(\xi),
\end{displaymath}
where $g \in \CO$ with $g\equiv 1$ on $Supp\ \hat{\psi}$, the functions $k_1, k_2, l_1$ (that can be matrix-valued) will
be short-range of order $O(\x^{-2})$. Observe that
there are two different types of correcting terms in this new
amplitude: on one hand the terms which contain $p(x,\xi+\lambda) \simeq
\Ps$ (when $\lambda \to \infty$) are correcting terms that ``live'' in
$\Hs^0$; on the other hand the terms which contain $\Pe$ are clearly
correcting terms that ``live'' in $\He^0$. Now, noting that
\begin{displaymath}
  B(x,\xi+\lambda) = -2(\xi+\lambda) \Pe + M(x,\xi+\lambda), \quad
  \quad M(x,\xi+\lambda) = -\frac{a^2(x)}{2(\xi+\lambda)}\Ga +
  a(x)\Gb,
\end{displaymath}
the amplitude $c(x,\xi,\lambda)$ can be written for $\xi \in Supp\ \hat{\psi}$ as
\begin{eqnarray}
  c(x,\xi,\lambda) & = & B(x, \xi + \lambda) P(x,\xi+ \lambda) -i \Ga
  \d_x P(x,\xi + \lambda), \nonumber \\
                   & = & B(x, \xi + \lambda) p(x,\xi+ \lambda) -i \Ga
  \d_x p(x,\xi + \lambda) \nonumber \\
                   &   & + \frac{1}{\lambda^2} \Big(
  -2(\xi+\lambda)\Pe k_1(x,\xi) + M(x,\xi+\lambda) \Pe k_1(x,\xi)
  - i\Ga \Pe \d_x k_1(x,\xi) \Big) \label{Step1} \\
                   &   & + \frac{1}{\lambda^2} \Big( B(x,\xi+\lambda)
  p(x,\xi+\lambda) l_1(x,\xi) -i\Ga (\d_x p(x,\xi+\lambda) l_1 +
  p(x,\xi+\lambda) \d_x l_1(x,\xi) )\Big) \nonumber \\
                   &   & + \frac{1}{\lambda^3} \Big(
  -2(\xi+\lambda)\Pe k_2(x,\xi) + M(x,\xi+\lambda) \Pe k_2(x,\xi)
  - i\Ga \Pe \d_x k_2(x,\xi) \Big).  \nonumber
\end{eqnarray}
Directly from the definitions we get the following asymptotics when
$\lambda \to \infty$
\begin{eqnarray*}
  \d_x p(x,\xi + \lambda) & = & +\frac{a'(x)}{2\lambda} \Gb \Ps -
  \frac{\xi a'(x)}{2\lambda^2} \Gb \Ps + O(\x^{-2} \lambda^{-3}), \\
  p(x,\xi+\lambda) & = & \Ps + O(\x^{-1} \lambda^{-1}), \\
  B(x,\xi+\lambda) p(x,\xi+\lambda) & = & O(\x^{-4} \lambda^{-3}).
\end{eqnarray*}
This leads to
\begin{eqnarray}
  c(x,\xi,\lambda) & = & -i\frac{a'(x)}{2\lambda} \Ga \Gb \Ps
  +i\frac{\xi a'(x)}{2\lambda^2} \Ga \Gb \Ps - \frac{2}{\lambda}
  \Pe k_1 - \frac{2\xi}{\lambda^2} \Pe k_1 \nonumber \\
                   &   &  + \frac{a(x)}{\lambda^2} \Gb \Pe k_1
  -i\frac{1}{\lambda^2} \Ga \Pe \d_x k_1 -i \frac{1}{\lambda^2} \Ga
  \Ps \d_x l_1 - \frac{2}{\lambda^2} \Pe k_2 \label{Step2}
  \\
                   &   & + O(\x^{-2} \lambda^{-3}). \nonumber
\end{eqnarray}
But the anticommutation properties of the Dirac matrices entail the
relations $\Ga \Gb \Ps = - \Pe \Gb$ and $\Gb \Pe = \Ps \Gb$. Hence the
amplitude $c(,\xi,\lambda)$ takes the form
\begin{eqnarray}
  c(x,\xi,\lambda) & = & \frac{1}{\lambda} \Pe \Big( i\frac{a'(x)}{2}
  \Gb - 2 k_1  \Big) \label{Step3}\\
                   &   & + \frac{1}{\lambda^2} \Pe \Big( -i\frac{\xi
                   a'(x)}{2} \Gb - 2\xi k_1 + i\d_x k_1 - 2k_2 \Big) +
  \frac{1}{\lambda^2} \Ps \Big( a(x) \Gb k_1 - i\d_x l_1 \Big)
  \nonumber \\
                   &   & + O(\x^{-2} \lambda^{-3}). \nonumber
\end{eqnarray}
Using (\ref{Step3}) we can cancel the terms of order less than
$O(\lambda^{-3})$ inductively. We choose
\begin{equation} \label{k1}
  k_1(x,\xi) = i\frac{a'(x)}{4} \Gb,
\end{equation}
which is obviously short-range of order $O(\x^{-2})$ and cancel the
first term in (\ref{Step3}). It remains then
\begin{eqnarray}
  c(x,\xi,\lambda) & = & \frac{1}{\lambda^2} \Pe \Big( -i\frac{\xi
                   a'(x)}{2}\Gb - i\frac{2\xi a'(x)}{4} \Gb -
                   \frac{a''(x)}{4} \Gb - 2 k_2 \Big)
                   \label{Step4} \\
                   &   & +\frac{1}{\lambda^2} \Ps \Big( i\frac{a'(x)
                   a(x)}{4}  -i \d_x l_1 \Big) + O(\x^{-2}
                   \lambda^{-3}), \nonumber
\end{eqnarray}
for $ \xi \in Supp\ \hat{\psi}$. We choose
\begin{equation} \label{l1}
  l_1(x,\xi) = \frac{a^2(x)}{8},
\end{equation}
and
\begin{equation} \label{k2}
  k_2(x,\xi) = -i\frac{\xi a'(x)}{2} \Gb - \frac{a''(x)}{8} \Gb,
\end{equation}

\par\noindent
The three correcting terms $k_1, l_1, k_2$ are short-range of order $O(\x^{-2})$ for $\xi$ in a compact set
and cancel the two first terms in (\ref{Step4}). This leads to
\begin{equation} \label{estimatesC}
\forall (x,\xi) \in \R^2, \forall \alpha,
  \beta \in \N, \quad |\d_x^\alpha \d_\xi^\beta c(x,\xi,\lambda)| \leq
  C_{\alpha\beta} \  \x^{-2-\alpha} \ \lambda^{-3}.
\end{equation}

\par\noindent
Eventually the new amplitude $P(x,\xi,\lambda)$ of $J^\pm(\lambda)$ is
defined as
\begin{displaymath}
  P(x,\xi,\lambda) = \left( p(x,\xi+\lambda) + \frac{1}{\lambda^2} \Big(
  i\frac{a'(x)}{4} \Pe \Gb + p(x,\xi+\lambda) \frac{a^2(x)}{8}
  \Big)\right.
\end{displaymath}
\begin{displaymath}
  \left. \ + \ \frac{1}{\lambda^3} \Big( -i\frac{\xi a'(x)}{2} \Pe \Gb -
  \frac{a''(x)}{8} \Pe \Gb \Big) \right) \ g(\xi),
\end{displaymath}
or equivalently
\begin{displaymath}
  P(x,\xi,\lambda) = \left( p(x,\xi+\lambda) + \frac{1}{\lambda^2} \Big(
  i\frac{a'(x)}{4} \Gb \Ps + p(x,\xi+\lambda) \frac{a^2(x)}{8} \Big)\right.
\end{displaymath}
\begin{equation} \label{FinalAmplitude}
  \left. + \ \frac{1}{\lambda^3} \Big( -i\frac{\xi a'(x)}{2} \Gb \Ps -
  \frac{a''(x)}{8} \Gb \Ps \Big) \right) \ g(\xi).
\end{equation}

\vspace{0,2cm}
\par\noindent
Mimicking the proof of Proposition 3.1, we have

\begin{lemma}
  For any $\psi \in \Hs^0$ such that $\hat{\psi} \in \CO$ and for $\lambda$ large, we have
  \begin{equation} \label{ApproximationWO}
    W^\pm(\lambda) \psi = \lim_{t \to \pm \infty} e^{itH(\lambda)}
    J^\pm(\lambda) e^{-it(D_x + \lambda)} \psi.
  \end{equation}
\end{lemma}

\vspace{0,2cm}
\par\noindent
Now, we can state the main result of this section:

\begin{lemma} \label{FundEst}
  When $\lambda$ tends to infinity, the following estimate holds:
  \begin{displaymath}
    \| (W^\pm(\lambda) - J^\pm(\lambda)) \psi \| = O(\lambda^{-3}).
  \end{displaymath}
\end{lemma}
\textbf{Proof}: Let us consider the case $(+)$. Thanks to (\ref{Property3}), we have to estimate
$\| D^+(t,\lambda) \psi \|$ where $D^+(t,\lambda)$ is the FIO with phase
\begin{displaymath}
\varphi^+ (x, \xi, \lambda,t) = x\xi + \frac{1}{2(\xi+\lambda)} \ \int_{x+t}^{+\infty} \ a^2 (s) \ ds
\end{displaymath}
(which is uniformly bounded with respect to $t$) and with amplitude $c^+ (x+t, \xi, \lambda)$.
\par\noindent
Let $\chi \in \CO$ be a cut-off function defined by $\chi(x)= 1$ if $\mid x\mid \leq \frac{1}{2}$,
$\chi (x) =0$ if $\mid x\mid \geq 1$ and let $\zeta = 1 - \chi$.
We have
\begin{equation} \label{Decomp}
\| D^+(t,\lambda) \psi \| \leq \| \chi ( { {2x} \over t})  D^+(t,\lambda) \psi \| +
\| \zeta ( { {2x} \over t})  D^+(t,\lambda) \psi \| \ := \ (1)+(2).
\end{equation}
First, we estimate the contribution $(1)$. On $Supp \  \chi ( { {2x} \over t})$,
 $\mid x+t\mid \geq {t \over 2}$. Using (\ref{estimatesC}) and the  continuity of FIOs, we have
\begin{equation} \label{Decomp1}
\| \chi ( { {2x} \over t})  D^+(t,\lambda) \| \ = \ O (<t>^{-2} \lambda^{-3}).
\end{equation}
Now, let us estimate  the contribution $(2)$. It is clear that
\begin{equation} \label{Decomp2}
(2) \  \ \leq \ \| \zeta ( { {2x} \over t})  D^+(t,\lambda) \zeta ( { {8x} \over t}) \psi \| \ + \
\| \zeta ( { {2x} \over t})  D^+(t,\lambda) \chi( { {8x} \over t}) \psi \| \  := \ (a)+(b).
\end{equation}
On $Supp \ \zeta ( { {8x} \over t})$, $\mid x\mid \geq \frac{t}{16}$.
Since $\psi \in {\mathcal{S}}(\R)$,  $\| \zeta ( { {8x} \over t}) \psi \| = O(<t>^{-N})$
 for all $N\geq0$. Thus, using the continuity of FIOs again, we obtain $(a)\ = \ O (<t>^{-N} \lambda^{-3})$.
 \par\noindent
 Now, we evaluate the contribution $(b)$ by using a standard non-stationary phase argument. We have
 \begin{equation} \label{Decomp2b}
\zeta ( { {2x} \over t})  D^+(t,\lambda) \chi( { {8x} \over t}) \psi (x) \ =\ (2\pi)^{-2}
\zeta ( { {2x} \over t}) \ \int_{\R^2} \ e^{i\Psi^+ (t,x,y,\xi,\lambda)} \ c (x+t,\xi,\lambda)\
\chi( { {8y} \over t}) \ \psi(y)\ d\xi\ dy,
\end{equation}
where
\begin{equation} \label{phasenonstat}
\Psi^+ (t,x,y,\xi,\lambda) = (x-y)\xi +\frac{1}{2(\xi+\lambda)} \ \int_{x+t}^{+\infty} \ a^2 (s) \ ds.
\end{equation}
In order to investigate possible critical point, we calculate
\begin{equation}\label{derivativephase}
\d_{\xi} \Psi^+ (t,x,y,\xi,\lambda) = x-y -\frac{1}{2(\xi+\lambda)^2} \ \int_{x+t}^{+\infty} \ a^2 (s) \ ds.
\end{equation}
On $Supp \ \zeta ( { {2x} \over t})$, $\mid x\mid \geq \frac{t}{4}$ and on
$Supp \ \chi ( { {8y} \over t})$, $\mid y\mid \leq \frac{t}{8}$. Moreover the integral which appears
 on the (RHS) of
(\ref{derivativephase}) is uniformly bounded with respect to $\lambda, x$ and $t$. Thus,
$\mid \d_{\xi} \Psi^+ (t,x,y,\xi,\lambda) \mid \geq C (1+t+\mid x\mid)$. We conclude by a standard argument of
non-stationary phase that for all $N \geq 0$, $(b) = O\ (<t>^{-N} \lambda^{-3})$.
Integrating over $(0,+\infty)$, we obtain Lemma \ref{FundEst}.
\\ $\diamondsuit$  \\


\subsection{High energy asymptotics of the scattering operator}

In this section we use the previous construction of the modifiers
$J^\pm(\lambda)$ to prove the reconstruction formula
(\ref{RF}). Recall that we want to find the
asymptotics as $\lambda \to \infty$ of the function (see
(\ref{ExpressionF}))
\begin{displaymath}
  F(\lambda) = < W^-(\lambda) \psi_1, W^+(\lambda) \psi_2 >,
\end{displaymath}
where $\hat{\psi_1}, \hat{\psi_2}$  have compact support. Using Lemma \ref{FundEst}, we get
\begin{displaymath}
  F(\lambda) = < J^-(\lambda) \psi_1, J^+(\lambda) \psi_2> + \
  O(\lambda^{-3}).
\end{displaymath}
Let us first give the asymptotic expansion of $J^\pm(\lambda)$ at
high energy. By the constructions above, we have for
$\lambda$ large
\begin{eqnarray}
  J^\pm(\lambda) \psi & = & \frac{1}{\sqrt{2\pi}} \int_\R e^{i \varphi^{\pm} (x,\xi,\lambda) }
  \Big( p(x,\xi+\lambda) + \frac{1}{\lambda^2} \big(
  i\frac{a'(x)}{4} \Gb + p(x,\xi+\lambda) \frac{a^2(x)}{8} \big)\label{Modifier} \\
                        &   & + \frac{1}{\lambda^3} \big( -i\frac{\xi a'(x)}{2} \Gb -
  \frac{a''(x)}{8} \Gb \big) \Big) \hat{\psi}(\xi)
  d\xi. \nonumber
\end{eqnarray}
Thus $J^\pm(\lambda)$ can be seen as pseudodifferential operators
whose symbols $j^\pm$ are given by
\begin{eqnarray*}
  j^\pm(x,\xi,\lambda) & = & e^{\frac{i}{2(\xi + \lambda)} \int_0^{\pm \infty} a^2(x+s) ds}
  \Big( p(x,\xi+\lambda) + \frac{1}{\lambda^2}
  \big(i\frac{a'(x)}{4} \Gb + p(x,\xi+\lambda) \frac{a^2(x)}{8} \big) \\
                       &   &
  + \frac{1}{\lambda^3} \big( -i\frac{\xi a'(x)}{2} \Gb -
  \frac{a''(x)}{8} \Gb \big) \Big).
\end{eqnarray*}
Hence, using Taylor expansion of $e^t$ at $t=0$ and using the
expansion
\begin{displaymath}
  p(x,\xi+\lambda) = \Big( 1 + \frac{a(x)}{2\lambda}\Gb - \frac{\xi
  a(x)}{2\lambda^2} \Gb \Big) \Ps + O(\lambda^{-3}),
\end{displaymath}
we get
\begin{eqnarray*}
  j^\pm(x,\xi,\lambda) & = & \Big(1 + \frac{i}{2\lambda} I^\pm -
  \frac{i\xi}{2\lambda^2} I^\pm - \frac{1}{8\lambda^2} \big( I^\pm
  \big)^2 + O(\lambda^{-3}) \Big) \\
                       &   & \qquad \qquad \Big( 1 +
  \frac{a(x)}{2\lambda}\Gb - \frac{\xi a(x)}{2\lambda^2} \Gb
  +i\frac{a'(x)}{4\lambda^2} \Gb + \frac{a^2(x)}{8\lambda^2} +
  O(\lambda^{-3}) \Big),
\end{eqnarray*}
where we denoted $I^\pm (x) = \int_0^{\pm\infty} a^2(x+s)ds$. If we
put together the terms with same order, we obtain
\begin{eqnarray*}
  j^\pm(x,\xi,\lambda) & = & 1 + \Big( \frac{iI^\pm}{2\lambda} +
  \frac{a(x)}{2\lambda} \Gb \Big) + \Big( -\frac{i\xi
  I^\pm}{2\lambda^2} - \frac{(I^\pm)^2}{8\lambda^2} +
  \frac{a^2(x)}{8\lambda^2} \Big) \\
                       &   & + \Big(\frac{ia(x)I^\pm}{4\lambda^2} -
  \frac{\xi a(x)}{2\lambda^2} + i\frac{a'(x)}{4\lambda^2} \Big) \Gb +
  O(\lambda^{-3}).
\end{eqnarray*}
We denote $R^\pm$ the differential operators
\begin{equation} \label{R}
  R^\pm = \Big(-\frac{iI^\pm D_x}{2} - \frac{(I^\pm)^2}{8} +
  \frac{a^2(x)}{2} \Big) + \Big(\frac{ia(x)I^\pm}{4} -\frac{a(x)
  D_x}{2} + i\frac{a'(x)}{4} \Big) \Gb,
\end{equation}
as a shorthand. Then the high energy asymptotics of $J^\pm(\lambda)$
are
\begin{equation} \label{AsymptoticModifier}
  J^\pm(\lambda) \psi = \psi + \frac{1}{\lambda} \Big(
  \frac{iI^\pm}{2} + \frac{a(x)}{2} \Gb \Big) \psi +
  \frac{1}{\lambda^2} R^\pm \psi + O(\lambda^{-3}).
\end{equation}

We now can prove Theorem \ref{ReconstructionFormula}. We have
\begin{eqnarray*}
  F(\lambda) & = & <J^-(\lambda) \psi_1, J^+(\lambda) \psi_2 > +
  O(\lambda^{-3}) \\
  & = & \Big< \Big(1 + \frac{iI^-}{2\lambda} + \frac{a(x)}{2\lambda}
  \Gb + \frac{1}{\lambda^2} R^- \Big) \psi_1, \Big(1 +
  \frac{iI^+}{2\lambda} + \frac{a(x)}{2\lambda} \Gb +
  \frac{1}{\lambda^2} R^+ \Big) \psi_2 \Big> \\
  & = & <\psi_1, \psi_2> + \Big< \Big( \frac{iI^-}{2\lambda} +
  \frac{a(x)}{2\lambda} \Gb \Big) \psi_1, \psi_2 \Big> + \Big< \psi_1,
  \Big( \frac{iI^+}{2\lambda} + \frac{a(x)}{2\lambda} \Gb \Big) \psi_2
  \Big>  \\
  &   & \quad + \frac{1}{\lambda^2} \Big( <R^- \psi_1, \psi_2> +
  <\psi_1, R^+ \psi_2> \Big) + \Big< \Big(\frac{iI^-}{2\lambda} +
  \frac{a(x)}{2\lambda} \Gb \Big) \psi_1, \Big( \frac{iI^+}{2\lambda}
  + \frac{a(x)}{2\lambda} \Gb \Big) \psi_2 \Big>, \\
  & =  & <\psi_1, \psi_2> + \frac{M}{\lambda} +
  \frac{N}{\lambda^2}.
\end{eqnarray*}
Let's compute the first order term $M$. Since $\Gb \psi_1 \in
\He^0$, the following terms vanish
\begin{equation} \label{Cancelation}
  <\Gb \psi_1,\psi_2> = <\psi_1, \Gb \psi_2> = 0.
\end{equation}
In consequence
\begin{displaymath}
  M = \ <\psi_1, \frac{i}{2}(I^+ - I^-) \psi_2> \ = \ \frac{i}{2} \int_\R
  a^2(s)ds <\psi_1, \psi_2>.
\end{displaymath}
But a direct calculation using (\ref{F}) and (\ref{RWImplicit}) shows
that $\displaystyle\int_\R a^2(s)ds = \frac{(l+\half)^2}{r_+}$. Whence
\begin{equation} \label{Order1}
  M = \frac{i(l+\half)^2}{2r_+} <\psi_1, \psi_2>.
\end{equation}
We compute now the second order term $N$. Notice first that by
(\ref{R}) and (\ref{Cancelation}), we get
\begin{displaymath}
  <R^\pm \psi_1, \psi_2> = \Big< \Big( -\frac{i I^\pm D_x}{2} -
  \frac{(I^\pm)^2}{8} - \frac{a^2(x)}{8} \Big) \psi_1, \psi_2 \Big>,
\end{displaymath}
and
\begin{displaymath}
  \Big< \Big(\frac{iI^-}{2} + \frac{a(x)}{2} \Gb \Big)
  \psi_1, \Big( \frac{iI^+}{2} + \frac{a(x)}{2} \Gb
  \Big) \psi_2 \Big> = <\psi_1, \frac{1}{4} I^- I^+ \psi_2 > +
  <\psi_1, \frac{a^2(x)}{4} \psi_2 >.
\end{displaymath}
Hence we obtain
\begin{displaymath}
  N =  -<\psi_1, \frac{i}{2} (I^+ - I^-) D_x \psi_2> + \frac{1}{8}
  <\psi_1, (I^+ - I^-)^2 \psi_2> + <\psi_1, \frac{a^2(x)}{2} \psi_2>.
\end{displaymath}
Using $\displaystyle\int_\R a^2(s)ds = \frac{(l+\half)^2}{r_+}$ again,
\begin{equation} \label{Order2}
  N =  - \frac{i(l+\half)^2}{2r_+} <\psi_1, D_x \psi_2> + <\psi_1,
  \frac{a^2(x)}{2} \psi_2>  + \frac{(l+\half)^4}{8 r_+^2} <\psi_1,
  \psi_2>.
\end{equation}
The two leading terms (\ref{Order1}) and (\ref{Order2}) give exactly
the reconstruction formula (\ref{RF}). Hence our main result is
proved. \\ $\diamondsuit$ \\
\vspace{0.5cm}

\noindent \emph{Acknowledgements}: The first author would like to
warmly thank Dietrich H\"afner (who suggested him this problem) for
his precious help.


\end{document}